\documentclass[10pt,twocolumn,floatfix]{iopart}
\usepackage{graphicx}
\usepackage{xcolor}
\usepackage[colorlinks=true, urlcolor=blue, citecolor=blue, linkcolor=blue]{hyperref}

\begin{document}

\title[Unconventional critical state in YBCO with a vortex-pin lattice]{\LARGE \textsf {Unconventional critical state in YBa$_{2}$Cu$_{3}$O$_{7-\delta}$ thin films with a vortex-pin lattice fabricated by masked He$^+$ ion beam irradiation}}

\author{\textsf{G~Zechner$^1$, K~L~Mletschnig$^1$, W~Lang$^1\footnote{Corresponding author}$, M~Dosmailov$^2$, M~A~Bodea$^2$ and J~D~Pedarnig$^2$}}
\address{$^1$ University of Vienna, Faculty of Physics, Electronic Properties of Materials, Boltzmanngasse 5, A-1090, Wien, Austria}
\address{$^2$ Johannes-Kepler-University Linz, Institute of Applied Physics, Altenbergerstrasse 69, A-4040 Linz, Austria}
\ead{wolfgang.lang@univie.ac.at}

\vspace{10pt}
\begin{indented}
\item[]This is an author-created, un-copyedited version of an article published in Superconductor Science and Technology {\bf 31} (2018) 044002 (8pp). IOP Publishing Ltd is not responsible for any errors or omissions in this version of the manuscript or any version derived from it.  The Version of Record is available online at \href{https://doi.org/10.1088/1361-6668/aaacc1}{https://doi.org/10.1088/1361-6668/aaacc1}.
\end{indented}

\begin{abstract}
Thin superconducting YBa$_{2}$Cu$_{3}$O$_{7-\delta}$ films are patterned with a vortex-pin lattice consisting of columnar defect regions (CDs) with 180~nm diameter and 300~nm spacing. They are fabricated by irradiation with 75~keV He$^+$ ions through a stencil mask. Peaks of the critical current reveal the commensurate trapping of vortices in domains near the edges of the sample. Upon ramping an external magnetic field, the positions of the critical current peaks are shifted from their equilibrium values to lower magnetic fields in virgin and to higher fields in field-saturated down-sweep curves, respectively. Based on previous theoretical predictions, this irreversibility is interpreted as a nonuniform, terrace-like critical state, in which individual domains are occupied by a constant number of vortices per pinning site. The magnetoresistance, probed at low current densities, is hysteretic and angle dependent and exhibits minima that correspond to the peaks of the critical current. The minima's positions scale with the component of the magnetic field parallel to the axes of the CDs, as long as the tilted vortices can be accommodated within the CDs. This behavior, different from unirradiated films, confirms that the CDs dominate the pinning.

\end{abstract}

%
\vspace{2pc}
\\
%
%
%
\ioptwocol

\section{Introduction}

Many materials, in their pure and clean form, are only marginally suitable for technical applications. It is by the introduction of controlled defects that superconductors can be tailored for many important properties, e.g., an enhanced critical current that is achieved by impeding the motion of the flux quanta of the magnetic field, called fluxons or vortices. Along these lines, many concepts have been explored both experimentally and theoretically.

Whereas enhanced vortex pinning in conventional superconductors has been achieved by metallurgical techniques, the brittle nature of cuprate superconductors precludes such fabrication methods. In high-$T_c$ superconductors (HTSCs), aligned columnar amorphous regions with diameters of a few times the coherence length have been introduced by irradiation with heavy ions and found to be a useful tool to enhance the critical current density \cite{CIVA97R}. However, the enhanced flux line pinning is strongly angle dependent, with a sharp maximum when the magnetic field is oriented parallel to the defect track \cite{HOLZ93,PROZ94}.

In contrast to such extended defects, point defects can be created by electron, proton, and light ion irradiation of HTSCs. For energies of up to few MeV the incident particles collide with a nucleus and displace it, eventually creating a collision cascade for high enough recoil energies. Since higher ion energies result in larger penetration depths but also in a reduced scattering cross section, a tradeoff between the possible thickness of the sample and a feasible ion fluence has to be found. Several studies \cite{SEFR01,LANG04} have revealed that He$^+$ ion irradiation of moderate energy is well suitable to tailor the superconducting properties of YBa$_2$Cu$_3$O$_{7-\delta}$ (YBCO) thin films by displacing mainly oxygen atoms, leading to a systematic reduction of $T_c$.

Recently, it has been demonstrated \cite{LANG06a,LANG09,PEDA10,SWIE12,TRAS13,HAAG14,ZECH17a} that the two concepts of columnar defects and point defects can be combined to create a regular array of cylindrical defect channels (CDs), in which superconductivity is locally suppressed, allowing to accommodate fluxons in a commensurate arrangement. Such vortex commensurability effects have been primarily studied in metallic superconductors \cite{MOSH10M} and the various patterns beautifully demonstrated by Lorentz microscopy \cite{HARA96b}. In HTSCs, however, due to their anisotropic layered structure, strong thermal fluctuations, $d$-wave gap symmetry, and large number of intrinsic defects such undertaking is more demanding.

Regular pinning landscapes are not primarily aimed at enhancing the critical current, but rather on many different ways of fluxon manipulation, like guided vortex motion \cite{WORD04,LAVI10}, vortex ratchets \cite{VILL03,SOUZ06,OOI07,PALA12}, and valves that are ingredients for low-dissipative computing applications \cite{HAST03,MILO07}. Such ideas, however, require that fluxons can be trapped in stable out-of-equilibrium positions that are defined by the artificial pinning landscape, and, ideally, can be probed by electrical quantities.

In a recent paper \cite{ZECH17a}, we have reported on the observation of hysteretic critical current and magnetoresistance in a YBCO film with an artificial, periodic pinning landscape and attributed this to two types of vortices, namely vortices trapped by extended defects and weaker pinned interstitial vortices. However, since it is known that YBCO films have strong intrinsic pinning \cite{DAM99} this notion calls for further support by additional experiments.

In this paper, we first investigate non-equilibrium vortex arrangements in a YBCO film with a columnar defect array (CDA), created by masked ion irradiation, starting from well-defined magnetization conditions. The virgin curve of the critical current as a function of the applied magnetic field is compared to down-sweep data, where the sample was put into an equilibrium state at various initial magnetic fields. In order to reveal whether pinning of vortices in the CDs or intrinsic pinning at interstitial positions is responsible for the observed hysteretic effects, the angle dependence of the resistance in tilted magnetic fields is examined.

\section{Experimental procedures}

Epitaxial thin films of YBa$_2$Cu$_3$O$_{7-\delta}$ are grown on (100) MgO single-crystal substrates by pulsed-laser deposition using 248~nm KrF-excimer-laser radiation at a fluence of 3.2~J/cm$^2$. The thickness of the film used in this work is $t_z = (210 \pm 10)$~nm. The critical temperature of the as-prepared film is $T_c \sim 90$~K, the transition width $\Delta T_c \sim 1$~K, and the self-field critical current density $j_c \sim 3$~MA/cm$^2$ at 77~K. For the electrical transport measurements a bridge with a length of $240~\mu \mathrm{m}$\ and a width $w = 60~\mu \mathrm{m}$\ is patterned by photolithography and wet chemical etching. The contacts are established on side arms of the sample in a four-probe geometry using sputtered Au pads with a voltage probe distance of $100\ \mu \mathrm{m}$.

An array of columnar defect cylinders is created in the YBCO film by masked ion beam irradiation (MIBS). The method is based on our previous observation \cite{LANG04} that  irradiation of YBCO with 75 keV He$^+$ ions leads to a reduction of the critical temperature $T_c$. At the applied fluence of $3 \times 10^{15}\ \mathrm{cm}^{-2}$ superconductivity is suppressed, leaving the irradiated regions normal conducting with a resistivity of $\rho \sim 1~\mathrm{m} \Omega$cm that is almost temperature independent between 20 and 310~K. This effect is due to the sensitivity of YBCO to displacements of the weakly-bound chain oxygen atoms \cite{WANG95b}, while the skeleton of YBCO's crystalline structure still remains intact.

\begin{figure}[t]
\centering
\includegraphics*[width=\columnwidth]{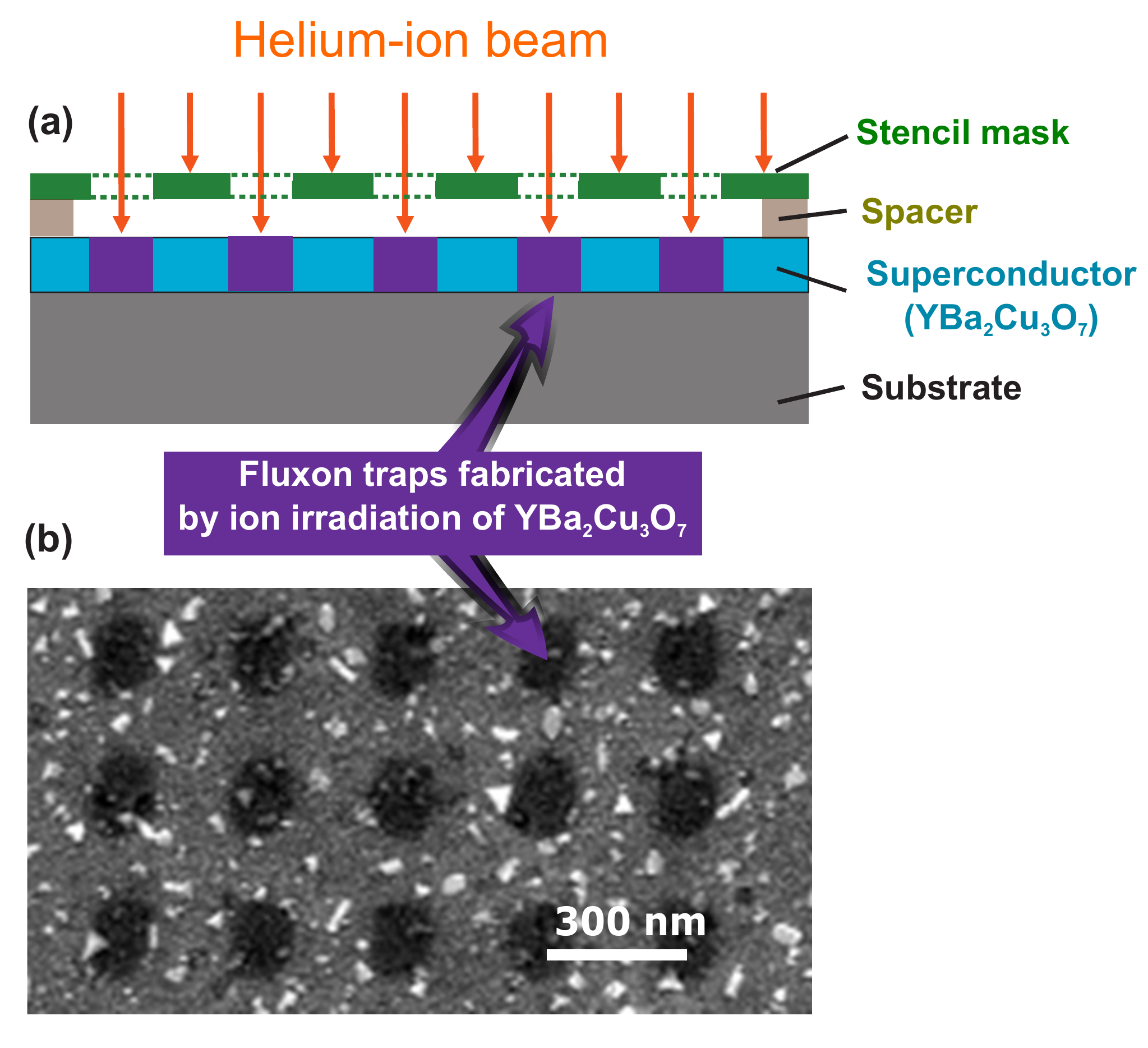}
\caption[]{(a) The principle of masked ion beam direct structuring (MIBS) to fabricate artificial fluxon traps in a superconducting thin film. (b) Scanning electron microscopy (secondary electron detection) picture of the surface of a thin YBCO film after MIBS with 75 keV He$^+$ ions. The dark regions correspond to an array of defect-rich, non-superconducting nano-cylinders.}
\label{fig:MIBS}
\end{figure}

The fabrication procedure is sketched in figure~\ref{fig:MIBS}(a). A thin Si stencil mask (custom fabricated with e-beam lithography by ims-chips, Germany) is mounted on top of the YBCO film and separated from its surface by a circumferential spacer layer of $1.5\ \mu$m-thick photoresist. By this procedure any possibly adverse contact between the surfaces of the mask and the YBCO film is avoided. The parallel alignment of the mask relative to the pre-patterned YBCO bridge was achieved by monitoring marker holes in the Si membrane with an optical microscope and was found to be better than 0.3$^\circ$. The mask features a square array of about $670 \times 270$\ holes with diameters $D = (180 \pm 5)$~nm and $d = (302 \pm 2)$~nm lattice constant, covering the entire YBCO bridge. MIBS provides a 1:1 projection of the hole pattern perforating the mask, since the ion beam can only reach the YBCO sample through these holes. The ion irradiation produces columnar defect-rich regions (CDs) in the sample \cite{LANG06a}. All other parts of the sample, as well as the electrical contacts, are protected from the irradiation.

The irradiation is performed in a commercial ion implanter (High Voltage Engineering Europa B. V.) with the beam oriented parallel to the sample's $c$-axis. To avoid any loss of oxygen by heating up the YBCO film during irradiation, the beam current is kept small and the sample stage is cooled. The beam is rapidly scanned over the sample stage area and the dose monitored by Faraday cups. Figure~\ref{fig:MIBS}(b) shows a scanning-electron microscopy picture of the columnar defect array (CDA), where dark areas represent the different escape rate of secondary electrons from the irradiated zones \cite{PEDA10}. The picture confirms that the irradiated areas are well-defined at the surface of the film and their diameter corresponds to $D = (180 \pm 5)$~nm of the mask, but simulations with SRIM \cite{SRIM} indicate that scattering of a few collision cascades away from the CD into regions of the sample that should be protected from irradiation might be the cause for the degraded $T_c \sim 47$~K in our sample. After storing the sample at room temperature for several weeks, $T_c$ increased by $\sim 2$~K due to partial annealing of some of the point defects. The slope of the resistivity vs temperature curve in the normal state remains unchanged after irradiation, indicating that the conducting channels are not oxygen depleted, while the room temperature resistivity increases by a factor $\sim 3$ \cite{HAAG14}. A similar reduction of $T_c$ has been observed in other studies of CDAs that were produced by masked ion irradiation and also attributed to the formation of stray point defects outside of the CD \cite{SWIE12}. In view of the above-mentioned observations it is useful to discuss the experiments in terms of $T/T_c$ rather than the absolute temperature $T$. Note that for the present investigations the spatial variation of the vortex pinning energy is a prerequisite and is realized by a modulation of the local critical temperature.

Magnetoresistance $R(B_a)$ and critical current $I_c(B_a)$ as a function of the applied magnetic field are measured in a closed-cycle cryocooler with in-field temperature control by a Cernox resistor \cite{HEIN98} to a stability of about 1 mK. The magnetic field is supplied by a revolvable electromagnet with a graduated scale of 1$^\circ$ resolution. In the angle-dependent magnetoresistance measurements, the magnitude of the magnetic field $B_a$ between the magnet's pole pieces is monitored by a calibrated Hall probe with an accuracy of $\pm 1$~mT, and, thus, is independent of the rotated position. The measurements are performed in constant Lorentz force geometry, i. e., the magnetic field is always perpendicular to the current direction. For the critical current measurements the magnetic field is applied perpendicular to the sample's surface and measured with a LakeShore 475 gaussmeter equipped with a HSE probe with a resolution of 0.1 $\mu$T, a zero offset $< 10 \mu$T, and a reading accuracy $<0.1 \%$. The current through the sample is generated by a Keithley 2400-LV constant-current source in both polarities to exclude thermoelectric signals and the voltages are measured by a Keithley 2182A nanovoltmeter.

Different procedures are employed to measure the critical current $I_c(B_a)$. The virgin curves are recorded after zero-field cooling (ZFC) the sample from 100~K into the superconducting state, with $B_a$ carefully adjusted to zero reading of the gaussmeter and partially compensating the earth's magnetic field.  Then, current is ramped up with exponentially increasing values until the voltage criterion of 100~nV, corresponding to $10 \mu$V/cm, is reached. Subsequently, $B_a$ is raised to the next value and the next $I_c(B_a)$ datum collected.

To compare a state with equilibrium vortex arrangements to the virgin curve, field-cooled (FC) measurements are conducted by setting $B_a$ to a predetermined value and cooling the sample in-field through the superconducting transition for \emph{every} data point. After a delay of 5 min to settle the target temperature, the $I_c(B_a)$ is measured. Subsequently, the sample is warmed to 100~K again, $B_a$ set to the next value, and the procedure repeated.

Down-ramped data, starting from an equilibrium and field-saturated state, are taken by initially applying a field $B_{FC}$ at 100~K and then cooling the sample into the superconducting state. $I_c(B_a)$ with $B_a \leq B_{FC}$ is measured while keeping the sample in the superconducting state.

\section{Results and discussion}

\begin{figure}[t]
\centering
\includegraphics*[width=\columnwidth]{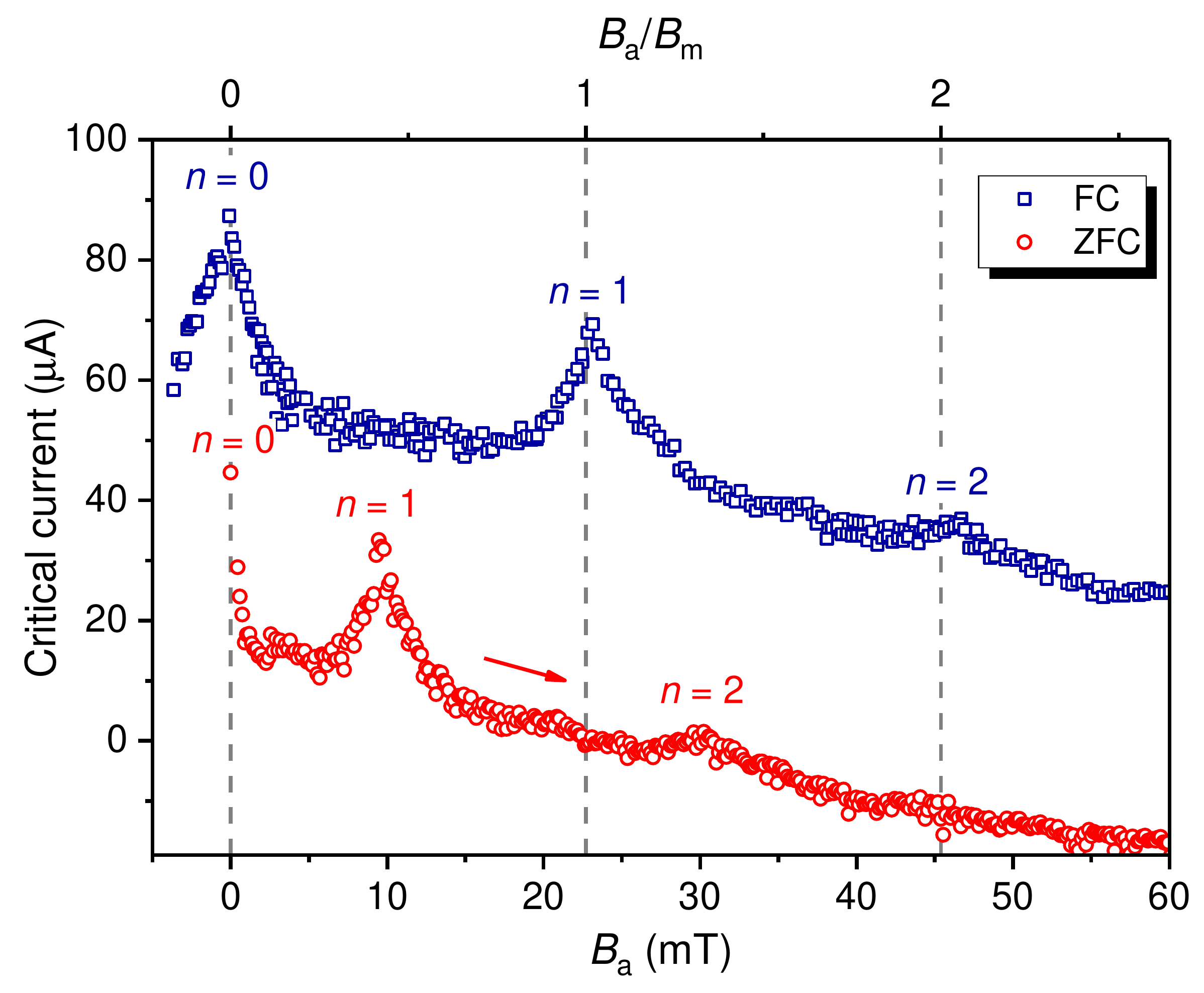}
\caption[]{Comparison of the critical current measured after FC from above $T_c$ to 38~K ($T/T_c = 0.78$) for every datum (blue squares) and the respective data after ZFC and ramping the applied magnetic field up (red circles). The curve is displaced by $-40~\mu$A for better visibility.}
\label{fig:Virgin}
\end{figure}

A fundamental property of the YBCO thin film with a CDA is illustrated in figure \ref{fig:Virgin} by a comparison of the critical current for FC data and the virgin curve after the sample has been ZFC and then the magnetic field ramped up.

The FC curve shows the peaks exactly at multiples of the matching field $B_m$
\begin{equation}
\label{eq:matching}
n B_m = n \frac{{{\phi _0}}}{{{d^2}}},
\end{equation}
where $\phi_0$ is the flux quantum and $n$ a rational number. For convenience we use $n = 0$ to denote the absence of vortices, $n<0$ for reversed vortex orientation, and $B_n$ for the position of peaks or minima that are related to matching conditions. Such commensurability between fluxons and artificial pinning arrays have been observed in several electrical transport experiments in both metallic \cite{FIOR78,LYKO93,METL94} and cuprate superconductors \cite{CAST97,OOI05,AVCI10,SWIE12,TRAS13,HAAG14,TRAS14,ZECH17a}.

\begin{figure}[h]
\centering
\includegraphics*[width=0.9\columnwidth]{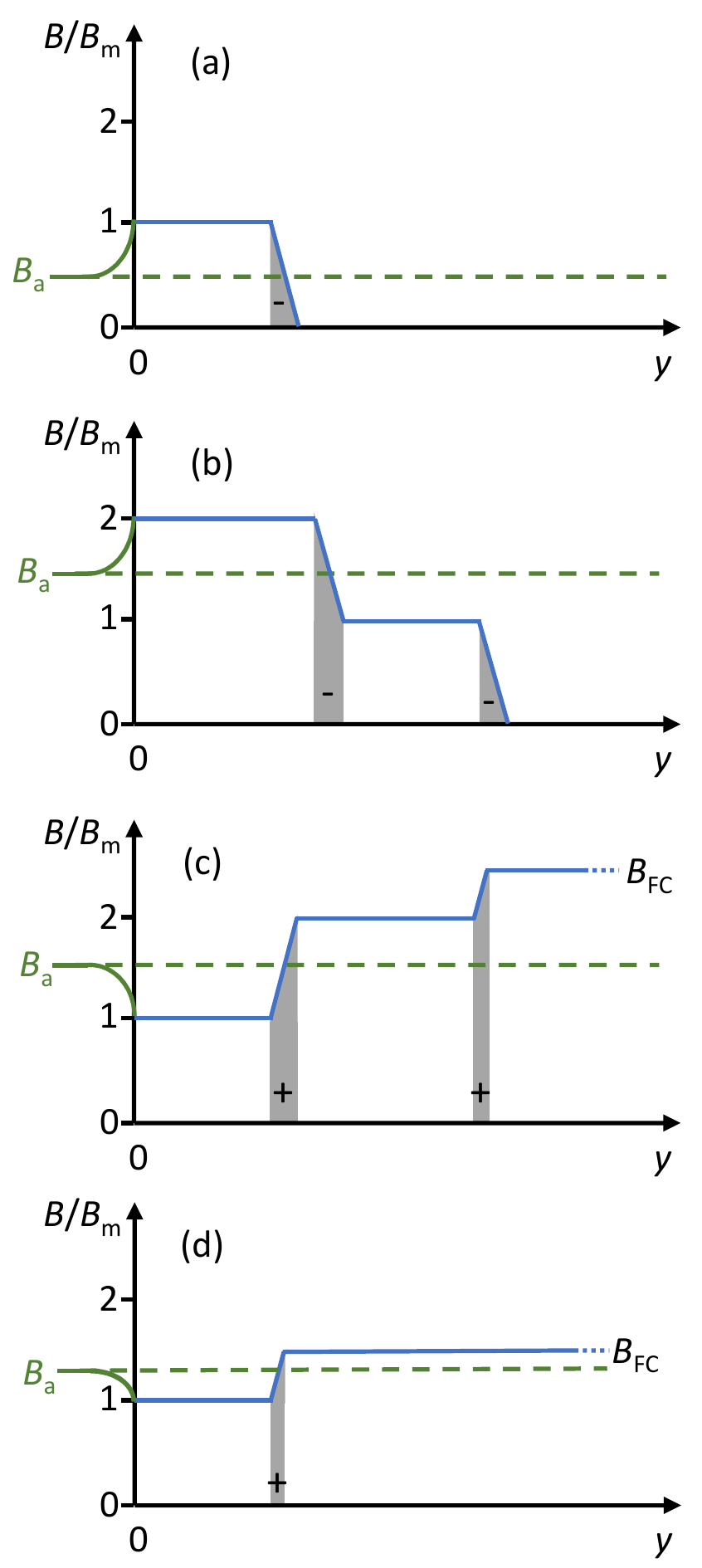}
\caption[]{Sketch of the proposed terrace-like critical state (not to scale), plotted from the edge of the sample along the width towards its center ($0 \leq y \leq w$). The local modulation of $B$ due to vortex cores is ignored.  The gray areas show the zones of the stratified shielding currents between the terraces and the sign marks their direction. The applied magnetic field $B_a$ is distorted due to demagnetization effects near the sample's edge. (a) Virgin curve at $n = 1$ matching ($B_a=B_1$). (b) Virgin curve at $n = 2$ matching ($B_a=B_2$). (c) $n = 1$ matching during down-ramping $B_a$ and starting from $B_{FC}=2.5 B_m$ (d) $n = 1$ matching during down-ramping $B_a$ and starting from $B_{FC}=1.5 B_m$}
\label{fig:terraces}
\end{figure}

The peaks of $I_c$ can be understood as a consequence of two different vortex pinning mechanisms. On one hand, $n$ fluxons can be trapped in the normal-conducting core of a CD. At the used irradiation fluence, the material inside a CD is converted to a normal-conducting, defect-rich metal with an almost temperature-independent resistivity of about 1~m$\Omega$cm \cite{LANG04}. Acting as a cylindrical antidot in the superconducting film it can accommodate many fluxons \cite{BUZD93,DORI00} due to the large ratio of the diameter of the CD to the in-plane coherence length $D / \xi_{ab}(0) \sim 130$, where $\xi_{ab}(0) \sim 1.4$~nm \cite{LANG95e}. A possible slight increase of $\xi_{ab}(0)$ due to the reduced $T_c$ in the irradiated YBCO would not change this consideration. These fluxons, pinned at CDs, are not moved by the transport current applied to the sample. Only by changing the applied magnetic field, the Lorentz force due to shielding current exceeds the pinning potential and the fluxons can hop between neighboring CDs \cite{SORE17}.

On the other hand, vortices at interstitial positions between the CDs are weaker pinned. In such a scenario with two different vortex pinning mechanisms, the critical current shows a peak when most available fluxons are trapped in the CDs due to commensurability effects. At the same time, the number of weaker pinned interstitial vortices that can give rise to dissipation is smallest. Such an interplay between vortices that are pinned at periodic defects and mobile interstitial vortices is also an important ingredient for the observation of vortex-ratchet effects in superconductors with asymmetric pinning landscapes \cite{VILL03,SOUZ06,OOI07,PALA12}.

In sharp contrast to our FC data, we observe a significantly different behavior in ZFC and magnetic-field ramped experiments, shown as red circles in figure \ref{fig:Virgin}. The peaks of the virgin curve after ZFC are displaced to lower field values which is an indication of both demagnetization effects in our thin YBCO films and strong irreversibility. Note that previous transport measurements in superconductors with regular pinning arrays showed no \cite{FIOR78,METL94,OOI05,AVCI10,SWIE12,TRAS13} or only slight \cite{LYKO93,CAST97} indications of irreversibility.

In a thin superconducting film, demagnetization effects provoke an overshoot of the external magnetic field at the edges of the sample leading to a breakdown of the Meissner state even in very small fields and to the penetration of vortices. In the ideal situation of an infinite thin film with a demagnetization factor of 1, the average internal field $\bar{B}$ equals the applied perpendicular field $B_a$ \cite{DORI08}. The large ratio $w/t_z = 286$ puts our sample close to this limit.

The critical current of the sample is determined by the motion of interstitial vortices in a small region near the edge of the sample \cite{BRAN93c} and does not lead to a redistribution of the strongly pinned fluxons in the CDs. The fact that the $n=1$ matching peak is observed already at $B_1 \sim 10\,\mathrm{mT} < B_m$ indicates a pronounced critical state with a domain of commensurable vortex pattern near the sample's edge \cite{ZECH17a} as it is outlined in figure~\ref{fig:terraces}(a). Upon further rising $B_a$, vortices propagate towards the interior of the sample and another domain at the edges forms, filled with two fluxons per CD resulting in $B_2 \sim 30\,\mathrm{mT} < 2 B_m$, establishing a terrace-like profile of the critical state, see figure~\ref{fig:terraces}(b).

Actually, for vortex-pin lattices, a terraced critical state has been theoretically predicted by Cooley and Grishin \cite{COOL95}. It is characterized by circumferential domains in the sample, inside which the pinning centers are occupied by the same number $n$ of fluxons and neighboring domains by $n \pm 1$. A similar domain structure has been revealed by a two-dimensional dynamic simulation of vortex arrangements in the critical state of superconductors with periodic pinning sites by Reichardt \etal \cite{REIC97} and inferred from scanning Hall probe measurements in a perforated Pb thin film by Silhanek \etal \cite{SILH11}. Another important consequence of such state is that the shielding currents are stratified into streamlines between the terraces and vanish inside, as indicated by gray areas in figure~\ref{fig:terraces}. Hence, the local current density in the streamlines is higher than its average over the sample's width.

\begin{figure}[t]
\centering
\includegraphics*[width=\columnwidth]{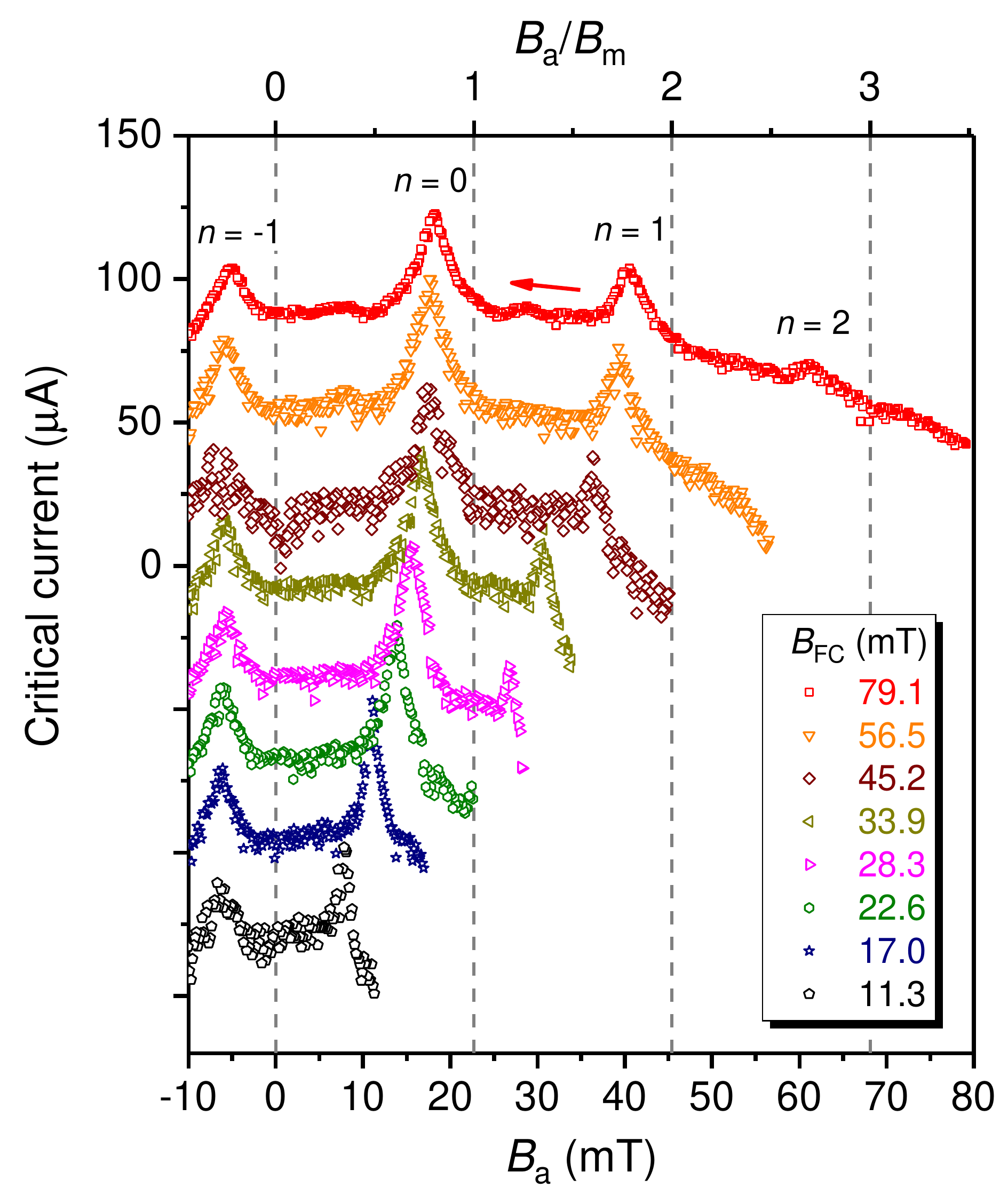}
\caption[]{Critical current measured during down-ramps of the applied magnetic field after establishing an equilibrium vortex arrangement by in-field cooling from 100~K to 38~K ($T/T_c = 0.78)$ at specified $B_{FC}$ values. The data at 79.1~mT are directly scaled, the other curves are displaced by $-30\;\mu$A each for better visibility.}
\label{fig:Saturat}
\end{figure}

\begin{figure}[t]
\centering
\includegraphics*[width=\columnwidth]{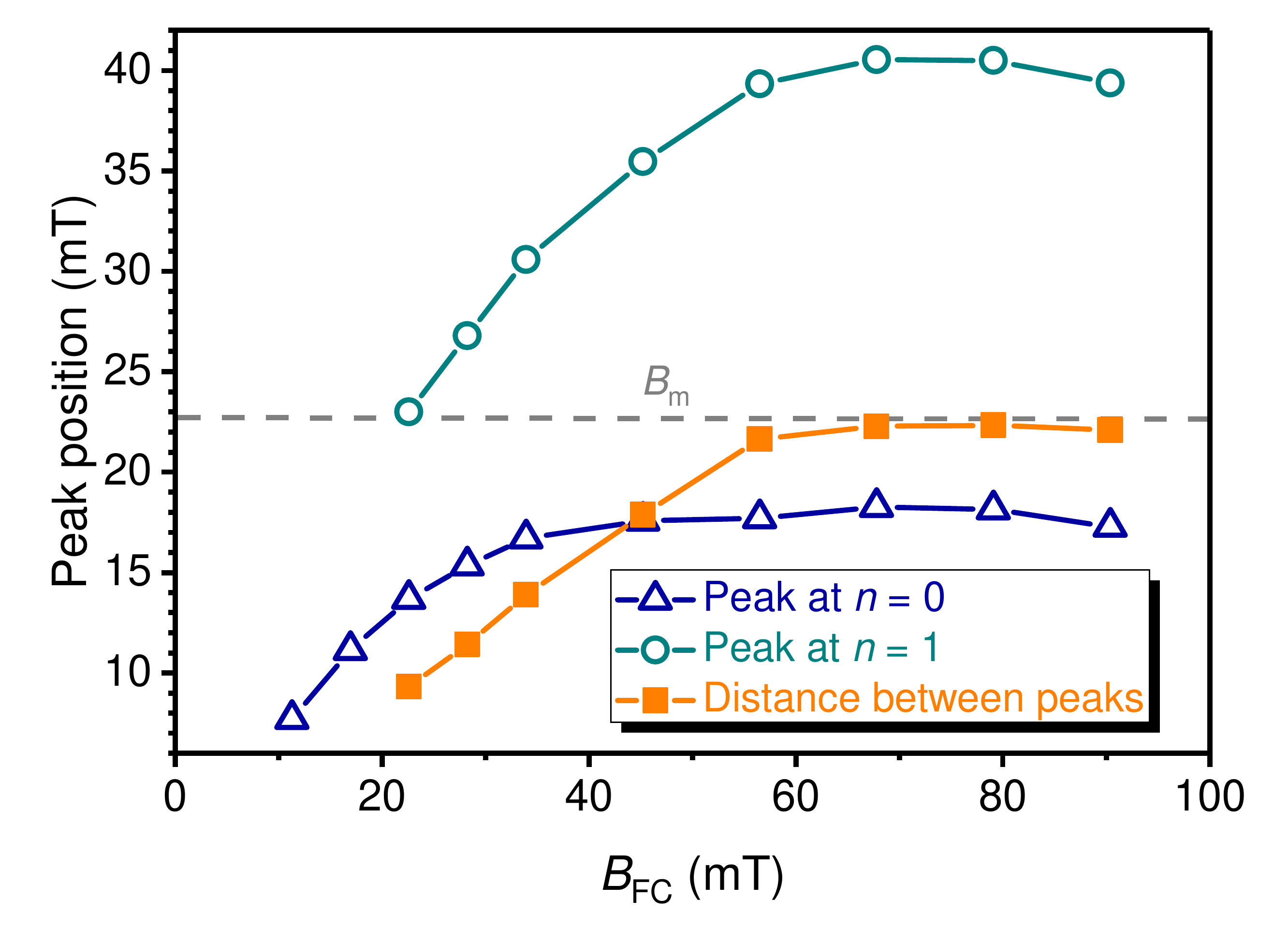}
\caption[]{Position of the $n= 0$ and $n = 1$ critical current peaks from figure~\ref{fig:Saturat} and their distance as a function of the initial field during cooling $B_{FC}$.}
\label{fig:Peaks}
\end{figure}

It is instructive to investigate the reverse situation, where a finite $B_{FC}$ is applied during cooling below $T_c$, thus allowing initially for an equilibrium arrangement of vortices and a field-saturated state within the sample. Field-cooling the sample with $B_{FC} > 80$~mT before $B_a$ is ramped down leads to an identical behavior as the one shown as the topmost curve in figure~\ref{fig:Saturat}, which, thus, can be seen as representative for higher $B_{FC}$. In all curves, the critical current peaks appear already at higher fields than predicted by Eq.~(\ref{eq:matching}). When $B_{FC}$ is reduced, the peaks are gradually shifted to lower $B_a$ values, except for the $n = -1$ peak. The peak positions are summarized in figure~\ref{fig:Peaks}, where it can be noticed that they remain stable for freezing fields $B_{FC} > 70\ \mathrm{mT} \sim 3 B_m$. Remarkably, the distance between all neighboring peaks at high enough $B_{FC}$ corresponds exactly to the matching field $B_m$.

Since the critical current is determined by vortex motion near the edge of the sample it probes the features of the outermost domain. Upon lowering $B_a$ the equilibrium state again transforms into a terraced critical state, where the terrace situated near the sample's edge has a local $B_T<B_a$ as depicted in figure~\ref{fig:terraces}(c). When it is populated by exactly one fluxon per CD, i.e., $B_T=B_m$, a peak in $I_c$ appears. Due to the fact that the local field $B_T$ in the outer terrace is lower than $B_a$, the peak is up-shifted from its equilibrium value of Eq.~(\ref{eq:matching}).

Strikingly, the distance between the peaks (orange squares in figure~\ref{fig:Peaks}) equals the matching field $B_m$ for high $B_{FC}$. According to Eq.~(\ref{eq:matching}) it indicates that exactly one flux quantum \emph{per CD} is removed over the \emph{entire} sample area, in line with the prediction of the model of a terraced critical state \cite{COOL95} and discussed in more detail in Ref.~\cite{ZECH17a}.

However, in two situations this condition is not met: The virgin curve in figure~\ref{fig:Virgin} demonstrates that initially the inner parts of the sample are not yet populated by fluxons, as sketched in figure~\ref{fig:terraces}(a) and (b). Thus, by increasing $B_a$ from $B_n$ to $B_{n+1}$, one additional fluxon is trapped only in CDs near the samples edges and the increase in average number of fluxons per CD over the entire sample is smaller than one, resulting in a compressed distance between peaks.

Conversely, by comparing figure~\ref{fig:terraces}(c) and (d) it is obvious that in the latter case the critical state profile near the samples's center is too flat to allow for extraction of appropriate number of fluxons. In this case, the peaks and their distance are both shifted to lower magnetic fields. A special situation happens at the $n = -1$ peak (after reversal of $B_a$). It corresponds to a virgin curve, displaced by a small field of opposite polarity still trapped in the center of the sample, since $|B_{-1}| < |B_{1}|$ of the virgin curve. However, the trapped field is similar whether it was established by ramping $B_a$ down from a high $B_{FC}$ or cooling the sample in an already low $B_{FC}$.

The existence of two different pinning mechanisms in our samples is an important ingredient to our reasoning. By an investigation of the angle dependence of the magnetoresistance in our sample we attempt to discriminate between pinning of vortices by CDs and the intrinsic pinning in the interstitial regions.

As mentioned before, the CDs are normal-conducting cylinders with a diameter $D=180~\mu$m and length $t_z=210$~nm. If the magnetic field is tilted off the CD's axis by an angle $\alpha$ its capability to accommodate fluxons remains essentially the same but the matching condition of Eq.~(\ref{eq:matching}) is modified by replacing $B_a$ with the component of $B_a$ parallel to the axis of the CD,

\begin{equation}
B_{\parallel} = B_a \cos{\alpha}.
\label{eq:scale1}
\end{equation}
However, when $\alpha$ exceeds $\alpha_{max} = \tan^{-1}(AR)$, where $AR = D/t_z = 0.82$ is the aspect ratio of the CDs, the pinning potential for fluxons is expected to decrease rapidly.

In unirradiated YBCO films, pinning of vortices is mainly caused by twinning and growth defects oriented parallel to the $c$ axis \cite{DAM99}. Angle-dependent measurements show a  maximum of the irreversibility temperature in a parallel field, extending to about $\alpha \sim 10^\circ$ at half maximum \cite{FIGU06}, but for larger $\alpha$ the irreversibility line follows the anisotropic superconductor model \cite{BRAN92}, where it is determined by an effective field,
\begin{equation}
B_{eff}=B_a \sqrt{\gamma^2 \sin^2 \alpha + \cos^2 \alpha},
\label{eq:scale2}
\end{equation}
with the anisotropy parameter $\gamma \sim 1/7$ in YBCO. Remarkably, pinning is stronger for $\alpha = 90^\circ$.

In our samples, however, additional point defects introduced by stray irradiation into the interstitial regions are present that favor elastic deformation of vortices and, thus, reduce the pinning potential of the twins {\cite{SEFR01}}. Thus, pinning effects around $\alpha = 0$ will be reduced and the overall pinning behavior be described by the effective-field approach of {Eq.~\ref{eq:scale2}}.

\begin{figure}[t]
\centering
\includegraphics*[width=\columnwidth]{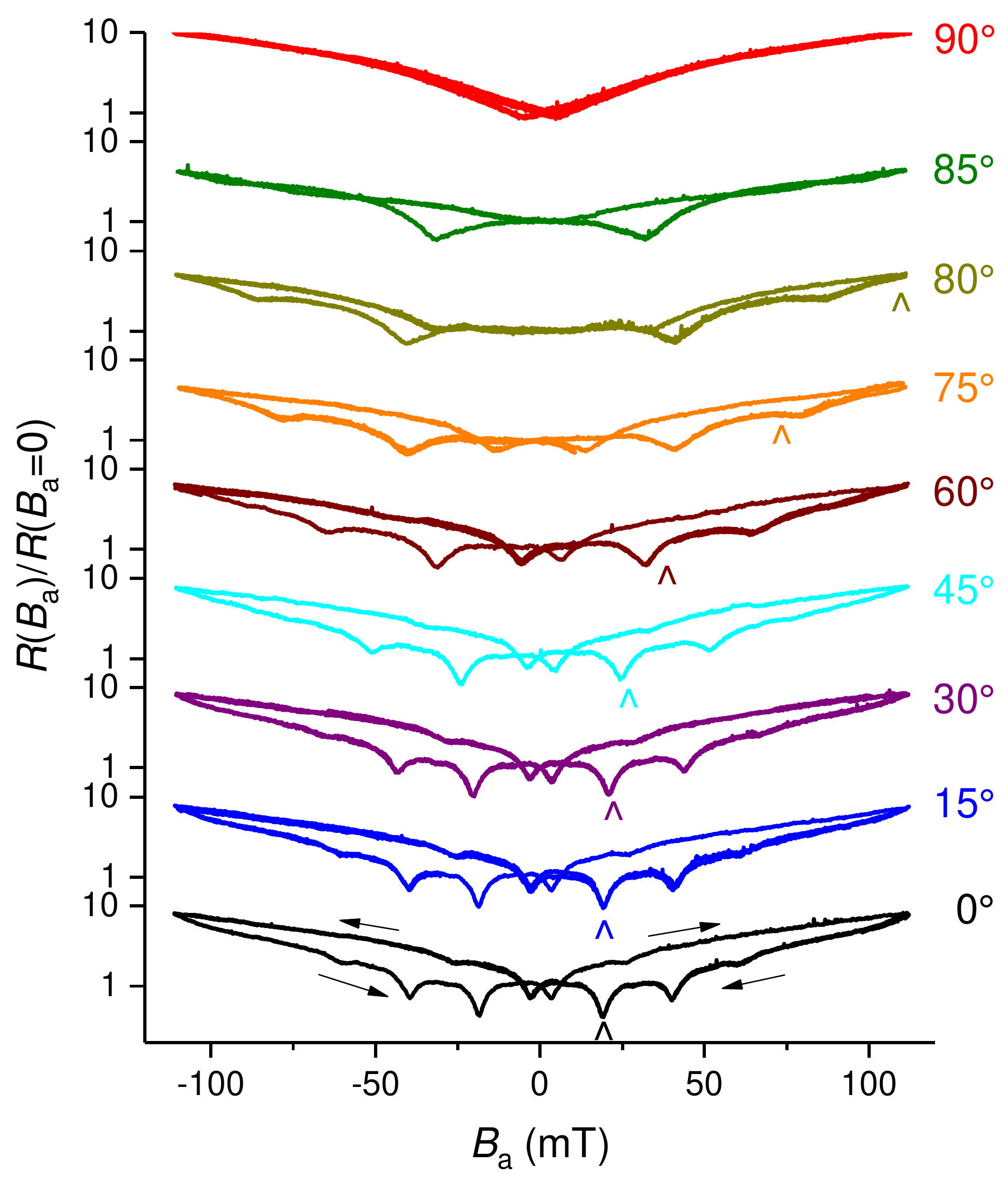}
\caption[]{Hysteretic behavior of the normalized resistance after zero-field cooling at several angles between the applied magnetic field and the crystallographic $c$ axis of YBCO at 36~K ($T/T_c = 0.77$). Data are recorded with a current of 200~$\mu$A, which is only about twice the typical critical current of the other measurements. The second sweep is shown to exclude the virgin curve, successive sweeps produce indentical traces. Arrows indicate the sweep direction and carets mark equal values of the component of $B_a$ that is parallel to the $c$ axis.}
\label{fig:angle}
\end{figure}

Figure~\ref{fig:angle} shows the normalized resistance of the YBCO thin film with the CDA as a function of the applied magnetic field $B_a$ after ZFC and ramping $B_a$ up and down. Data are shown for the second cycle and exclude the virgin curve. The two most important features are, on the one hand, pronounced minima of the resistance that can be best seen in the down-sweep curves. These are a signature of reduced vortex mobility at commensurate vortex arrangements and are complementary to the above-discussed peaks of $I_c$. Since magnetoresistance is probed at low currents of about $2 I_c$, data from both kinds of measurements can be compared. Minima in the resistance and peaks of $I_c$ occur at the same $B_a$ \cite{ZECH17a}.

In analogy to the topmost curve in figure~\ref{fig:Saturat}, the distance between two adjacent minima is $\Delta B = B_1 - B_0 \sim B_m = 22.7$~mT at $\alpha = 0^\circ$. On the other hand, all curves, except the one at $\alpha = 90^\circ$, where $B_a$ is parallel to the CuO$_2$ planes of YBCO, exhibit a pronounced hysteresis that is another manifestation of the unconventional critical state in YBCO films with a regular pinning array \cite{ZECH17a}. The $n = 0$ minimum, marked by a caret in figure~\ref{fig:angle}, occurs at $B_0 = 19$~mT at $\alpha = 0^\circ$. When the magnetic field is moderately tilted away from the sample's $c$ axis, the qualitative picture does not change much, but the minima are displaced to higher $B_a$ values.

\begin{figure}[t]
\centering
\includegraphics*[width=\columnwidth]{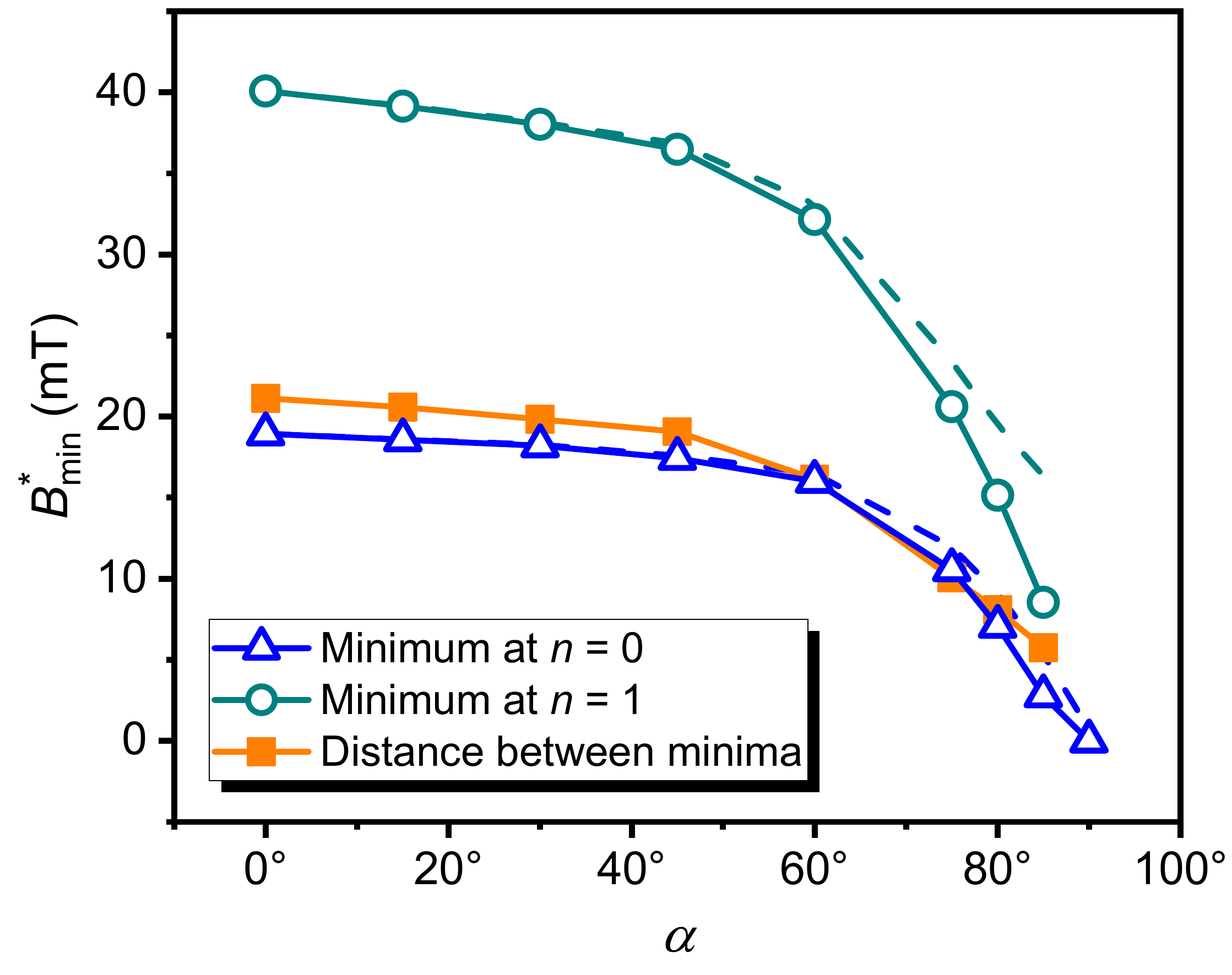}
\caption[]{Position of the minima observed in figure~\ref{fig:angle} as a function of the magnetic field's tilt angle $\alpha$ with respect to the $c$ axis of YBCO. Data are scaled to the magnetic field component parallel to the $c$ axis according to Eq.~{(\ref{eq:scale1})} (symbols and full lines) and, for comparison, to the effective field of Eq.~{(\ref{eq:scale2})} (broken lines). Note that constant {$B^*_{min}$} would indicate conformance to the theoretical models.}
\label{fig:minima}
\end{figure}

In figure~\ref{fig:minima} the positions of the resistance minima are scaled to the magnetic field's component parallel to the axes of the CDs, $B_{\parallel}$ of Eq.~(\ref{eq:scale1}). For comparison, scaling to Eq.~(\ref{eq:scale2}) is displayed, too, but is only marginally different. Almost constant values of $B^*_{min}$ indicate a valid scaling, but for $\alpha > 45^{\circ}$ a significant drop can be noticed. The maximum tilt angle of a fluxon within a CD is $\alpha_{max} \sim 40^{\circ}$ and apparently, if exceeded, results in a reduced pinning potential of the CDs.

At higher tilt angles $\alpha > \alpha_{max}$, the unconventional critical state gradually breaks down and $B_0$, $B_1$, and $\Delta B$ decrease below their respective values of $B_{\parallel}$. Finally, at $\alpha = 90^{\circ}$, vortices are intrinsically pinned between the CuO$_2$ layers \cite{TACH89,FEIN90} and irreversibility line measurements in YBCO films indicate stronger pinning for $\alpha = 90^\circ$ \cite{FIGU06}. While the resistance in our samples at magnetic fields above the observed matching effects, $B = 100$~mT, decreases from $R_{\alpha=0^\circ}= 48.8\ \mathrm{m}\Omega$ to $R_{\alpha=90^\circ}= 8.7\ \mathrm{m}\Omega$, the hysteresis in the $R(B_a)$ curves becomes much smaller.

From the observation that angle-dependent pinning at the CDA follows the expected scaling according to Eq.~(\ref{eq:scale1}) as long as $\alpha < \alpha_{max}$, the absence of a narrow pinning peak around $\alpha=0^\circ$ due to pinning at twin boundaries, and the fact that the hysteresis is almost vanishing in the strong intrinsic pinning situation at $\alpha=90^\circ$ in YBCO we conclude that the pronounced hysteresis of $R(B_a)$ is caused by pinning of vortices by the CDA and not by a pinning mechanism specific to the unirradiated parts of the sample.

\section{Conclusions}

In a thin YBCO film with a square array of columnar defects produced by masked ion irradiation an unconventional critical state emerges from the coexistence of two pinning mechanisms: strong pinning at the artificial defects and weaker intrinsic pinning in the unirradiated areas. This critical state is characterised by peaks in the critical current and minima in the resistance, respectively, with remarkable properties. If the sample is field-saturated at high fields and the field then ramped down, the $I_c$ peaks associated with a commensurate occupation of the columnar defects by fluxons appear at different fields than in the virgin curves and at a distance that corresponds exactly to the matching field. A previous theoretical prediction of a terraced critical state, consisting of domains in which each defect is populated by the same number of fluxons, can explain such behavior. These commensurability effects of a non-equilibrium vortex arrangement depend on the component of a tilted magnetic field that is parallel to the axes of the defect columns but gradually vanish when the angle exceeds a critical value defined by the geometry of the columns. This finding identifies the trapping of fluxons in the defect array as the main pinning mechanism.

\section{Acknowledgments}
We appreciate the help of Klaus Haselgr\"ubler with the ion implanter. This article is based upon work from COST Action CA16218 (NANOCOHYBRI), supported by COST (European Cooperation in Science and Technology). M.D. acknowledges the European Erasmus Mundus (Target II) program for financial support.

\section*{References}
\bibliographystyle{iopart-num}

\end{document}